\documentstyle[12pt]{article}
\def\tenbf{\normalsize\bf}
\def\tenrm{\normalsize\rm}
\def\tenit{\normalsize\it}
\def\elevenbf{\normalsize\bf}
\def\elevenrm{\normalsize\rm}

\def\vereq#1#2{\lower3pt\vbox{\baselineskip1.5pt \lineskip1.5pt
\ialign{$\m@th#1\hfill##\hfil$\crcr#2\crcr\sim\crcr}}}

\def\eslt{E\llap/_T}
\def\Re{{\cal R \mskip-4mu \lower.1ex \hbox{\it e}}\,}
\def\Im{{\cal I \mskip-5mu \lower.1ex \hbox{\it m}}\,}

\def\tg{\tilde g}

\def\tq{\tilde q}

\def\tw{\widetilde W}
\def\tz{\widetilde Z}

\renewenvironment{thebibliography}[1]
 { \elevenrm
   \begin{list}{\arabic{enumi}.}
    {\usecounter{enumi} \setlength{\parsep}{0pt}
     \setlength{\itemsep}{3pt} \settowidth{\labelwidth}{#1.}
     \sloppy
    }}{\end{list}}
\begin{document}
\begin{flushright}
UH-511-773-93 \\
FSU-HEP-930829 \\
August 1993
\end{flushright}
\begin{center}{{\tenbf DETECTING TOP SQUARKS AT THE TEVATRON}
\vglue 1.0cm
%%%
%{\ninerm (For 20\% Reduction to 8.5 $\times$ 6 in Trim Size)\\}
%\vglue 1.0cm
{\tenrm Howard Baer$^1$, John Sender$^2$ and Xerxes Tata$^2$ \\}
\vglue 0.5cm
\baselineskip=13pt
{\tenit $^1$Department of Physics, Florida State University, Tallahassee, FL
32306\\}
\vglue 0.3cm
{\tenit $^2$Department of Physics and Astronomy, University of Hawaii,\\
 Honolulu, HI, 96822\\}
\vglue 0.8cm
{\tenrm ABSTRACT}}
\end{center}
\vglue 0.3cm
{\rightskip=3pc
 \leftskip=3pc
 \tenrm\baselineskip=12pt
 \noindent
We study the signal from the pair production of $t$-squarks at the
Tevatron under the assumption that their two-body decay to charginos as well as
their three-body decay to $W$ bosons is kinematically forbidden.
In this case, the stop dominantly decays via $\widetilde{t_1}\rightarrow
c\tz_1$,
so that the signal consists of two charm jets together with $\eslt$.
We reevaluate this signal using ISAJET~7.01, and show that
if the stop mass is below about 100 GeV, there
should be as many as 50-70 events in the accumulated data sample
of the CDF and D0 experiments even if the LSP is as heavy as
50 GeV. We have also studied the possibility of tagging the $c$-jet
by its semileptonic $\mu$ decay, but find that the event rates
are too small for this to be viable except for values of stop and
LSP masses that yield a robust signal via the conventional $\eslt$
search.
\vglue 0.6cm}

{\elevenbf\noindent 1. Introduction\hfil}

The spectacular success of the recent Tevatron runs has enabled
experimentalists
to search for signals from a variety of extensions of the Standard Model (SM)
including new gauge bosons, compositeness, exotic quarks, leptoquarks and,
of primary interest to us, supersymmetric particles. The CDF
Collaboration\cite{CDF},
based on an analysis of 4.3 $pb^{-1}$ of data have published limits of
around 100 GeV (160 GeV if $m_{\tq}=m_{\tg}$) on the masses of squarks and
gluinos of the minimal supersymmetric model (MSSM).  By now, the D0 and CDF
experiments at the Fermilab Tevatron have, between them, accumulated an
integrated luminosity of almost 40 $pb^{-1}$\cite{CORNELL}, and are expected to
collect a data sample in excess of 100 $pb^{-1}$ by the end of the current run.
It is, therefore, reasonable to ask whether the large anticipated data sample
opens up the possibility of detecting other sparticles at the Tevatron.

It has recently been pointed out\cite{SIGNALS} that the multilepton signals
from the cascade decays of gluinos and squarks, which provide indirect
evidence for the existence of charginos($\tw_{i}$) and neutralinos($\tz_{i}$),
should make it possible to extend
the Tevatron search for gluinos and squarks to 250-300 GeV for the favourable
case $m_{\tq} \simeq m_{\tg}$. It has also been shown\cite{BT,ASPECTS} that
with an accumulated data sample of about 100 $pb^{-1}$,
the trilepton signal from
the continuum $\tw_{1}\tz_{2}$ production\cite{AN} would enable the CDF and D0
experiments to extend the direct search for charginos and neutralinos to
regions of MSSM parameter space beyond the range of LEP experiments.
Continuing our study of other SUSY signals that may be accessible at
the Tevatron, we focus here on the strategies for detection of the
$t$-squark which can be considerably lighter\cite{LIGHT} than all the other
squarks even in supergravity models where all the sfermions have a common mass
at an ultrahigh energy unification scale.\\

{\elevenbf\noindent 2. Why is the top squark different?\hfil}

Supersymmetry must be a broken symmetry. Since the dynamics of SUSY breaking is
as yet unknown, the
breaking of supersymmetry is parametrized by a rather large number of
soft SUSY-breaking parameters constrained only by SM gauge invariance. The
proliferation
of parameters can be reduced by making further assumptions about the
symmetries of the dynamics of SUSY breaking. Within the simplest
supergravity GUTS\cite{SUGRA}, SUSY breaking can be parametrized
by a common scalar mass, a common gaugino mass together with a susy-breaking
trilinear scalar coupling at unification scale. These masses and couplings
are then evolved down to the weak scale, leading to the familiar
relation ``GUT relation'' between the three gaugino masses. Since the
first two generations of squarks and sleptons dominantly interact via
gauge interactions, their masses evolve in the same way. As a result,
these squarks (sleptons) are essentially degenerate, with squarks heavier
than sleptons on account of their QCD interactions.

The masses of third generation squarks ($\widetilde{t_L}, \widetilde{t_R}$
and
$\widetilde{b_L}$) are not expected to
conform to these simple patterns because of their large
Yukawa interactions. These reduce the diagonal masses in comparision to those
of
the other squarks, in much the same way that they drive a Higgs scalar
mass squared to negative values, resulting in the radiative
breaking\cite{SUGRA}
of electroweak symmetry. Furthermore, the Yukawa interactions also induce
$\tilde{f_L}$-$\tilde{f_R}$ mixing terms proportional to the corresponding
{\it fermion} mass, and so, are most important for the t-squarks. These
off-diagonal terms in the $t$-squark mass matrix split the top squark masses,
reducing the mass of the lighter stop ($\widetilde{t_1}$) state even further.
In
fact, $m_{\widetilde{t_1}}$ may be as light as 50 GeV even if the other squarks
and gluinos have masses of several hundred GeV. This led a group of
us\cite{GODBOLE} to study the phenomenology of a light $\widetilde{t_1}$ and
its
effect on the CDF top quark search at the Tevatron. It had been concluded
that a $\widetilde{t_1}$ with a mass just beyond the LEP limit could well have
escaped detection in the analysis of the 1991 data sample. In this
study we improve on this parton level study and
reevaluate the stop signal using ISAJET 7.01/ISASUSY 1.0\cite{BPPT} with
a view to assess the prospects for stop detection during the current
Tevatron run.\\

{\elevenbf\noindent 3. Top-squark decay patterns.\hfil}

The decay patterns of the top squark have been discussed at length in
Ref.\cite{GODBOLE} and will only be briefly reviewed here.
If $\widetilde{t_1}$ is heavier than the the chargino, the tree-level two-body
decay
$\widetilde{t_1}\rightarrow b\tw_{1}$ dominates; in this case, stop pair
production
is signalled by $n-leptons + m-jet$ events ($n=1$ or 2) so that top quark
pair production (whose cross section is an order of magnitude larger) is
a formidable background.
If $m_{\tw{_1}}+m_b \geq m_{\tilde{t_1}}\geq M_W+m_b+m_{\tz{_1}}$,
the decay $\widetilde{t_1}\rightarrow bW\tz_{1}$ dominates, and the situation
is similar to that above. If this decay is kinematically forbidden
(as is the case for stops in the mass range of interest), and
the charginos as well as the sneutrinos are heavier than $\widetilde{t_1}$,
the flavour changing one loop decay $\widetilde{t_1} \rightarrow c+\tz_{1}$
dominates\cite{HIKASA,GODBOLE} the tree level four-body decays of the stop.
In the remainder of our
analysis, we assume that the branching fraction for this two body loop
decay is 100\%.\\

{\elevenbf\noindent 4. Stop signals at the Tevatron.\hfil}

Although $t$-squark production is not yet included in ISAJET, we can
simulate the signals for stop pair production by generating
$\widetilde{b_R}$-pair events (which have the same production
cross section as stop pairs) and using the FORCE command to decay the
$\widetilde{b_R}$ into a $c$-quark and an LSP. In our computation, we have
used the set I structure functions of Eichten {\it et. al.}\cite{EHLQ}.
The production cross section is about twice as large as shown in
Ref.\cite{GODBOLE}. This is partly due to a difference in structure functions,
but, more importantly, due to a difference in the scale used in the evaluation
of $\alpha_s$. Initial and final state parton showers, charm quark
fragmentation and decay and the underlying event structure are also
incorporated.
Aside from QCD radiation, the stop signal then consists of two $c$ jets
together with $\eslt$ from the escaping LSP's, and without $c$ tagging, is
identical to the signals from squark production, where the squark directly
decays to the LSP.

We have modelled the experimental
conditions at the Tevatron by incorporating a toy calorimeter with segmentation
$\Delta\eta\times\Delta\phi = 0.1\times 0.09$ and extending to $|\eta | = 4$
into our simulation. We have assumed an energy resolution of $70\% /\sqrt{E_T}$
($15\% /\sqrt{E_T}$) for the hadronic (electromagnetic) calorimeter. Jets are
defined to be hadron clusters with $E_T > 15$ GeV in a cone of $\Delta R =
\sqrt{\Delta\eta^2 +\Delta\phi^2} = 0.7$.
We have also incorporated the following cuts and triggers:

\noindent ({\it i}) We require that the jets lie within
$|\eta_j|$ $\leq 3.5$. All jets are required to be separated by at
least $30^{\rm o}$ in azimuth from $\vec{\eslt}$.

\noindent ({\it ii}) We require $n_j \geq 2$, with at least one jet within
$|\eta_j \leq 1|$.

\noindent ({\it iii}) If $n_j = 2$, we further require
$\Delta\phi$($j_1$,$j_2$) $\leq 150^{\rm o}$.

\noindent ({\it iv}) We veto events containing leptons (from the $c$-jet) with
$p_T$($l$)$\geq 20$ GeV to reduce the background from $W\rightarrow l\nu$
($l=e$ or $\mu$).

\noindent ({\it v}) We have required $\eslt \geq 50$ GeV\cite{CDF} to reduce
backgrounds from QCD heavy flavours and mismeasured jets.

Aside from heavy flavour QCD and jet mismeasurement backgrounds, SM sources
of $n \geq 2$ $ jets + \eslt$ events include, ({\it a})
$Z\rightarrow\nu\bar{\nu}$
production,
({\it b}) $W\rightarrow\tau\nu$ production (where the hadronically
decaying $\tau$ is assumed to be one of the jets), and ({\it c}) $W\rightarrow
l\nu$ ($l = e$ or $\mu$), where the jets are due to
initial state QCD radiation. We have not attempted to quantify the QCD
backgrounds which are expected to be detector-dependent, but have been guided
in our thinking by earlier experimental analyses\cite{CDF} which suggest that
the $Z$ and $W$ backgrounds just mentioned are indeed the dominant SM sources
of
these jet events with $\eslt \geq 50$ GeV.

The $\eslt$ distributions from the SM backgrounds ({\it a-c}) is shown in
Fig. 1 along with the corresponding distribution from stop pair production
for two representative choices of parameters where the stop decays
via $\widetilde{t_1}\rightarrow c\tz_1$: (A) $m_{\widetilde{t_1}} = 85$ GeV,
$m_{\tz_1} = 20$ GeV, and (B)  $m_{\widetilde{t_1}} = 125$ GeV,
$m_{\tz_1} = 40$ GeV. The distributions shown here are {\it before}
any cuts have been applied. We see that the $\eslt$ cut ({\it v}) designed
for reducing the QCD backgrounds also served to remove much of the $W$ and
$Z$ backgrounds.
\begin{figure}
\vspace{3.0in}
\caption{The normalized $\eslt$ distribution from top squark pair production
and background processes discussed in the text at the Tevatron collider.}
\end{figure}

We have studied several distributions that might help to further distinguish
the stop signal from background. The $p_T$ distribution of the fastest jet in
the signal events for the two cases introduced above, as well as for the SM
backgrounds ({\it a-c}) {\it after} the cuts ({\it i-v}) is shown in Fig. 2. As
expected, the fast jets from the three background sources (which arise
from QCD radiation) are
softer than the corresponding jets from the signal -- but for the $\eslt$
requirement, the background distribution would have been backed up against
the $p_{Tmin}$($j$) cut of 15 GeV. In contrast, we
see that the bulk of the signal events include a jet with $p_T\geq 50$ GeV
even for the lighter stop case in the figure.
\begin{figure}
\vspace{3.0in}
\caption{The normalized $p_T$($fast$-$jet$) distribution from top squark pair
production and related backgrounds at the Tevatron collider, after
cuts discussed in the text.}
\end{figure}

The azimuthal separation between the $\vec{\eslt}$ and the nearest jet
is shown in Fig. 3 for the stop signals and the $W$ and $Z$ backgrounds
discussed above. We see that the signal distribution is approximately
flat as may be expected from the fact that $\vec{\eslt}$ is made up
of two independently produced LSPs. In contrast, since the jets in
the background are recoiling against the produced $W$ or $Z$, the distribution
tends to peak at large angular separation. We see that requiring
$\Delta\phi \leq 90^{\rm o}$ significantly increases the signal to
background ratio.
\begin{figure}
\vspace{3.0in}
\caption{The azimuthal angle separation between $\vec{\eslt}$ and closest jet
from top squark pair production
and from SM background processes at the Tevatron collider,
after cuts.}
\end{figure}

Shown in Fig. 4 are contours of fixed signal cross sections in the
$m_{\widetilde{t_1}}-m_{LSP}$ plane including the cuts ({\it i-v})
together with ({\it a}) $\Delta\phi$($j$,$\vec{\eslt}$)$\leq 90^{\rm o}$,
and ({\it b}) if $\Delta\phi$($j$,$\vec{\eslt}$)$\leq 90^{\rm o}$,
$p_T$($j_{fast}$)$\geq 50$ GeV; $p_T$($j_{fast}$)$\geq 80$ GeV otherwise.
The cuts in Fig. 4{\it b} retain more of the signal for heavier stops
without letting in background events. Also shown on the figure
are the background levels expected from $W$ and $Z$ events at the Tevatron.
Several comments are worthy of note:

\noindent ({\it i}) The $W\rightarrow\tau$ background shown in the figure
assumes
that the $\tau$, if it decays hadronically, can give rise to one of the
two jets. Since $\tau$ jets almost always have a charged multiplicity
of 1 or 3, we believe that it should be possible to discriminate these
from the signal jets with high efficiency. Our purpose in showing these
cross sections is to allow the reader to assess the $\tau-jet$ discrimination
that is necessary to be able to see the $t$-squark signal.
We see that a discrimination of 1:10 is ample for this purpose.

\noindent ({\it ii}) The background from $Z\rightarrow \nu\bar{\nu}$ production
can be
subtracted since it should be possible to directly measure high $p_T$
$Z$ decays to leptons and use the branching ratios measured at LEP.

\noindent ({\it iii}) The cross section (after cuts) for a stop with a mass
of up to 100 GeV exceeds 2 $pb$ for an LSP as heavy as
50 GeV. Thus in excess 50-70 stop events may already be present
in the collective data sample of the CDF and D0 experiments; the
corresponding background from $Z$ + $jets$ production yields a comparable
number of events. If heavy flavour and QCD backgrounds are indeed
negligible, this would correspond to a 6-7$\sigma$ effect. This
conclusion should, however, be viewed in proper perspective since we have not
included any non-physics backgrounds in our analysis.

\noindent ({\it iv}) It is interesting to see that we find an observable signal
even
when the LSP is relatively heavy. This differs significantly from the
conclusion of the parton level calculation of
Ref.\cite{GODBOLE} where it was concluded that the signal
would be unobservable for LSP masses much larger than 20 GeV. This
conclusion was traced to the fact that for large values of $m_{\tz_1}$, the
two LSPs soaked up much of the energy so that
the charm partons became too soft to pass the cuts. In our present calculation,
the $\widetilde{t_1}\bar{\widetilde{t_1}}$ pair can be produced with
substantial
$p_T(\widetilde{t_1}\bar{\widetilde{t_1}})$; the resulting final state charm
jets can hence have
substantial $p_T$ even if $m_{\tz_1}$ is large.
\begin{figure}
\vspace{6.0in}
\caption{Contour plots in (pb) of top squark signal cross-sections after cuts,
and associated background rates.}
\end{figure}

Up to this point, we have made no use of the fact that the signal always
contains $c$-quark jets. It is clear that SM backgrounds would be considerably
reduced if it were to be possible to tag at least one of the $c$-quarks.
This led us to consider the possibility of using a muon from the semi-leptonic
decay of one of the $c$ quarks as a tag. The signal would then consist of $\mu
+ n_j \geq 2 + \eslt \geq 50$ GeV, where the muon is within a cone of
$\Delta R = 0.4$ about one of the jets. We require that $p_T$($\mu$)$ \geq 3$
GeV for the muon to be identifiable. We further require that
either  $\Delta\phi$($j$,$\vec{\eslt}$)$\leq 90^{\rm o}$ {\it or}
$p_T$($j_{fast}$)$\geq 50$ GeV. The signal cross section contours, with these
cuts, are shown in Fig. 5 together with our background estimates from
$W\rightarrow\tau\nu\rightarrow\mu\nu\nu\nu$, $W\rightarrow\mu\nu$
and $Z\rightarrow\nu\bar{\nu}+c\bar{c}$ or $b\bar{b}$ processes. To estimate
these, we have generated 140K (130K) $W$ ($Z$) events of each type,
and find 8,5 and 6 events, respectively,
pass our cuts. It should be kept in mind that ISAJET does not include
the full $2 \rightarrow 3$ matrix elements for $Zc\bar{c}$ or $Zb\bar{b}$
production;
in our simulation, these events come from radiation of initial state
gluons followed
by splitting into $b\bar b$ or $c\bar c$ pairs.
\begin{figure}
\vspace{3.0in}
\caption{Contour plots in (pb) of top squark signal cross-sections containing
an identifiable muon, after cuts, and associated background rates.}
\end{figure}

Our conclusions from Fig. 5 are
pessimistic. Even with an integrated luminosity of 100 $pb^{-1}$,
there are just 5-10 tagged signal events for $m_{\widetilde{t_1}}$ and
$m_{\tz_1}$ values where the $\eslt$ signal in Fig. 4 might be difficult
to observe above the $Z\rightarrow\nu\bar{\nu}$ background. Since the sum of SM
backgrounds to the signal have comparable cross sections, we believe that
it will be difficult to distinguish the stop signal over statistical
fluctuations of
the backgrounds. We further observe that the 0.2 $pb$ (this is the
sum total of the three backgrounds) contour in Fig. 5 roughly tracks
the contour with a cross section roughly equal to that from the $Z$ backround
in Fig. 4. We, therefore, infer
that the use of muon tagging
to extend the region where the stop signal might be observable in the $\eslt$
sample even with
several hundred $pb^{-1}$ of integrated luminosity does not appear
viable. Observation of tagged $\eslt$ events would, however, be important
since it indicates that the signal might be due to the production of
$\tilde{b},\tilde{c}$ or, of course, $\tilde{t}$ squarks. We also remark that
it
would be worth investigating whether vertex tagging can be used to tag
the charm jets in stop pair
events to increase the signal to background.

{\elevenbf\noindent 5. Summary and outlook\hfil}

Motivated by the fact that the $t$-squark may be considerably lighter
than other squarks, we have reinvestigated its signals at the Fermilab
Tevatron using ISAJET~7.01 under the assumption that it decays via
$\widetilde{t_1}\rightarrow c\tz_1$. We have shown that in the $\eslt$
data sample that has already been accumulated by the CDF and D0
experiments, there may well be 50-100, or more, stop events that
should, after suitable cuts, be identifiable above $W$ and $Z$ backgrounds
(these have previously\cite{CDF} been shown to dominate those from QCD) even
if the LSP is relatively heavy (see Fig. 4).
We also studied the possibility of tagging the charm jet via its semi-leptonic
decay. While this led to an observable rate for stop events for values
of parameters where the signal was observable by the conventional $\eslt$
search (this would, within the MSSM context, indicate the production
of a $c$-, $b$- or $t$-squark), the tagged signal (see Fig. 5) was found to be
too small for larger values of stop and LSP masses.

\vglue 0.5cm

{\elevenbf\noindent 6. Acknowledgements\hfil}
\vglue 0.4cm

We are grateful to the members of the LSGNA collaboration for participating
in the free exchange of many ideas during the Workshop. We especially thank
J.~Hewett and T.~Rizzo for their initiative which made this possible. This
research was supported in part by the US Dept. of Energy Contract No.
DE-FG05-87ER40319 and DE-AM03-76SF00235.

\vglue 0.5cm
{\elevenbf\noindent 7. References \hfil}
\vglue 0.4cm

\end{document}